\documentclass[12pt,twoside,a4paper]{article}
\usepackage[dvips]{epsfig}
\def\lsim{\:\raisebox{-0.5ex}{$\stackrel{\textstyle<}{\sim}$}\:}
\begin{document}
\thispagestyle{empty} 
\title{
\vskip-3cm
{\baselineskip14pt
\centerline{\normalsize DESY 08--109 \hfill ISSN 0418--9833}
\centerline{\normalsize MZ-TH/09--15 \hfill} 
\centerline{\normalsize May 2009 \hfill}} 
\vskip1.5cm
{\bf Inclusive $D^{*}$-meson production \\
in $ep$ scattering at low $Q^2$ \\
in the GM-VFN scheme at NLO}
\author{G.~Kramer$^1$ and H.~Spiesberger$^2$ \vspace{2mm} \\
{\normalsize $^1$ II. Institut f\"ur Theoretische
  Physik, Universit\"at Hamburg,}\\ 
\normalsize{Luruper Chaussee 149, D-22761 Hamburg, Germany} \vspace{2mm}\\
\normalsize{$^2$ Institut f\"ur Physik,
  Johannes-Gutenberg-Universit\"at,}\\ 
\normalsize{Staudinger Weg 7, D-55099 Mainz, Germany} \vspace{2mm} \\
}} 
\date{}
\maketitle
\begin{abstract}
\medskip
\noindent
We have calculated the next-to-leading order cross sections for the
inclusive production of $D^{*}$-mesons in $ep$ collisions at HERA for 
finite, although very small $Q^2$. In this $Q^2$-range, the same 
approximations as for photoproduction can be used. Our calculation 
is performed in the general-mass variable-flavour-number scheme. In 
this approach, large logarithms of the charm transverse momentum are
resummed and finite terms depending on $m^2/p_T^2$ are kept in the hard 
scattering cross sections. The theoretical results are compared with 
recent data from the ZEUS collaboration at HERA. On average, we find 
good agreement.
\end{abstract}

\clearpage

\section{Introduction}

The understanding of the dynamics of charm quark production at HERA 
has been improved considerably over the last ten years by the H1 and 
ZEUS collaborations, who have performed many measurements of inclusive 
$D^{*}$-meson production in the photoproduction mode of $ep$ 
collisions with almost vanishing virtuality ($Q^2 \simeq 0$) of the 
exchanged photon, as well as in the deep-inelastic (DIS) mode (with 
photon virtuality $Q^2 > 0$). The theoretical description of heavy 
quark production in the framework of perturbative QCD is complicated 
due to the presence of several large scales appearing in this process. 
In DIS, $Q^2$ is large, but also the transverse momentum $p_T$ of the 
produced $D^{*}$-meson may be large. In addition, depending on the 
kinematic range considered, also the mass $m$ of the charm quark 
may have to be taken into account. Different calculational schemes 
have been developed which can be applied for an interpretation of 
experimental data, depending on the specific kinematical region and 
the relative importance of these three scales.  

In the case of relatively small transverse momentum, $p_T \lsim m$, 
the fixed-flavour number scheme (FFNS) is usually applied. Here one 
assumes that the light quarks ($u$, $d$, $s$) and the 
gluon are the only active flavours within the proton and the photon. 
In this scheme cross sections for $e+p \to e'+D^{*}+X$ have been 
calculated for DIS in Ref.\ \cite{1} and for photoproduction in Ref.\ 
\cite{2}. In photoproduction, where $Q^2 \simeq 0$, the direct 
process has to be supplemented with the resolved process, where the 
photon acts as a source of partons which interact with partons in the 
proton. These two interaction modes are needed to describe the 
singular region at $Q^2 = 0$ for massless quarks appropriately. In the
FFN scheme \cite{1,2} the charm quark appears only in the final state 
of the direct and resolved processes, via the hard scattering of light 
partons, including the photon. The charm quark mass $m$ is explicitly 
taken into account together with the transverse momentum of the 
produced $D^{*}$-meson; this approach is therefore expected to be 
reliable when $p_T$ and $m$ are of the same order of magnitude. In 
the FFNS, the charm quark mass acts as a cutoff for the initial- and 
final-state collinear singularities and sets the scale for the 
perturbative calculation. The mass $m$ is fully retained in the 
calculation of the hard-scattering cross sections.

In the complementary kinematical region where $p_T \gg m$, calculations
are usually based on the zero-mass variable-flavour-number scheme
(ZM-VFNS). This is the conventional parton model approach where the
zero-mass parton approximation is applied also to the charm quark,
although its mass is not small and large compared with $\Lambda_{QCD}$.
In the ZM-VFNS, the charm quark acts also as an incoming parton with its
own parton distribution function (PDF) in the proton and in the photon
leading to additional direct and resolved contributions. Usually, charm
quark PDFs and also the fragmentation functions (FFs), describing the
transition $c \to D^{*}$, are defined at an initial scale $\mu_0$ chosen
equal to the charm mass $m$. Then this is the only place, where the
charm mass enters in this scheme.  The $D^{*}$-meson is produced by
fragmentation from the charm quark produced in the hard scattering
process; but also fragmentation from the light quarks and the gluon
has to be taken into account.  The well-known factorization theorem
provides a unique procedure for incorporating the FFs into the lowest
order (LO) and next-to-leading-order (NLO) perturbative calculations.
The predictions obtained in this scheme are expected to be reliable only
in the region of large $p_T$ since all terms of the order $m^2/p_T^2$
are neglected in the hard scattering cross section. Calculations for
$D^{*}$-production in the ZM-VFNS have been performed some time ago for
photoproduction in Ref.\ \cite{3} and for DIS in Ref.\ \cite{4}.

A unified scheme that combines the virtues of the FFNS and the ZM-VFNS
is the so-called general-mass variable-flavour-number scheme (GM-VFNS)
\cite{5}. In this approach the large logarithms $\ln(p_T^2/m^2)$, which
appear due to the collinear mass singularities in the initial and final
state, are factorized into the PDFs and FFs and summed by the well known
DGLAP evolution equations \cite{6} for the PDFs and FFs. The
factorization is performed following the usual $\overline{\rm MS}$
prescription which guarantees the universality of both PDFs and FFs. At
the same time, mass-dependent power corrections are retained in the
hard-scattering cross sections, as in the FFNS. It is expected that this
scheme is valid not only in the region $p_T^2 \gg m^2$, but also in the
kinematic region where $p_T$ is larger than a few times the charm mass
$m$ only. We should emphasize that in the GM-VFN scheme, the
incorporation of the fragmentation $c \to D^{*}$ is based on the
factorization theorem; this is a prerequisite for applying the FFs in
different processes. In the usual FFNS calculation, this is not the case
and non-perturbative FFs for the transition $c \to D^{*}$ can be
supplemented on purely phenomenological grounds only.

It is the purpose of this work to present theoretical results for the
$D^{*}$-production cross section in $e^{\pm}p$ scattering with finite
non-zero photon virtuality in the region $0.05 < Q^2 < 0.7$ GeV$^2$ and
discuss a comparison with experimental results obtained with the ZEUS
detector at HERA \cite{7}. We shall calculate all cross sections with
the same kinematical constraints as in the ZEUS analysis \cite{7} in the
GM-VFNS. Since in the ZEUS experiment the photon virtuality is small,
the application of the photoproduction approximation is justified where
the $Q^2$-dependence in the hard scattering cross sections is neglected.

Details of our calculation have been described recently in Ref.\ 
\cite{8}. In this work, we had also studied theoretical uncertainties  
of the photoproduction cross section in the GM-VFNS due to various 
possible choices for input variables, as for example, the proton and 
photon PDFs and the $D^{*}$ FFs. In that reference one can also find 
a discussion of the dependence on the factorization and 
renormalization scales and the influence of the charm quark mass $m$. 
The application of the theoretical framework described there to the 
present case of low-$Q^2$ DIS is straightforward and amounts to an 
adjustment of the parameters entering the Weizs\"acker-Williams 
approximation \cite{9} for the flux of the virtual photon. Instead of 
$Q_{\rm min}^2 = m_e^2 y^2 /(1-y)$ and the value for $Q_{\rm max}^2$ 
given by the anti-tagging condition of the final electron (positron) 
as used in the measurement of the photoproduction process, we have to 
fix $Q_{\rm min}^2$ and $Q_{\rm max}^2$ to the values used in the ZEUS 
analysis. We will consider the kinematic range as in Ref.\ \cite{7}, 
i.e.\ $0.05 < Q^2 < 0.7$ GeV$^2$.

The outline of the paper is as follows. In Section 2 we give a short 
description of the various input options used in the calculation. 
Section 3 contains our results and the comparison with the 
experimental data from ZEUS.


\section{Input choices for the calculation}

The $D^{*}$-electroproduction cross section $\sigma_{ep}(\sqrt{s})$ 
at the $ep$ center-of-mass energy $\sqrt{s}$ is related to the 
photoproduction cross section at center-of-mass energy $W_{\gamma p}$, 
$\sigma_{\gamma p}(W_{\gamma p})$, in the following way:
\begin{eqnarray}
\sigma_{ep}(\sqrt{s}) 
= 
\int_{y_{\rm min}}^{y_{\rm max}} dy
f_{e\gamma}(y,Q_{\rm min}^2,Q_{\rm max}^2)\sigma_{\gamma p}(y\sqrt{s})
\, .
\label{gpcrosss}
\end{eqnarray}
Here, $f_{e\gamma}$ is the energy spectrum of the exchanged virtual 
photon which in the Weizs\"acker-Williams approximation \cite{9} is 
given by
\begin{equation}
f_{e\gamma}(y,Q_{\rm min}^2,Q_{\rm max}^2) 
=
\frac{\alpha}{2\pi} 
\left[\frac{1+(1-y)^2}{y}
\ln \frac{Q_{\rm max}^2}{Q_{\rm min}^2} 
+ 2 m_e y \left(\frac{1}{Q_{\rm max}^2}-\frac{1}{Q_{\rm min}^2}\right)
\right] \, .
\label{fegwwa}
\end{equation}
The photon flux $f_{e\gamma}$ depends on $y$, $Q_{\rm min}^2$ and 
$Q_{\rm max}^2$. The range of $y$, $y_{\rm min} \leq y \leq 
y_{\rm max}$, as well as the limits $Q_{\rm min}^2$ and 
$Q_{\rm max}^2$, are determined by the cuts and bin limits in the 
experimental analysis. In photoproduction, $Q_{\rm min}^2 \propto 
m_e^2$ is very small, whereas in the ZEUS analysis $Q_{\rm min}^2 = 
0.05$ GeV$^2$ and $Q_{\rm max}^2 = 0.7$ GeV$^2$, or within these 
limits for the measurement of the $Q^2$ distribution. $\alpha$ is 
the electromagnetic fine structure constant and $y = E_{\gamma}/E_e$, 
the ratio of the energies of the incoming photon and electron, is 
determined by the inelasticity $y = Q^2/(2P \cdot q)$ where $P$ and 
$q$ are the 4-momenta of the incoming proton and the photon.

The cross section for direct photoproduction in Eq.\ (\ref{gpcrosss}) is
a convolution of the proton PDF, the fragmentation function for the
transition of parton $a$ to the observed $D^{*}$-meson (where $a = u$,
$\bar{u}$, $d$, $\bar{d}$, $s$, $\bar{s}$, $c$, $\bar{c}$, and $g$) and
the cross section for the hard scattering process $\gamma b \to aX$. For
the resolved contribution, an additional convolution with the photon
PDFs has to be performed. The hard scattering cross sections are
calculated including next-to-leading order corrections of the order
$O(\alpha_s)$. The PDFs are evolved at NLO. For the photon PDF we use
the set GRV92 of Ref.\ \cite{10}, converted to the $\overline{\rm MS}$
factorization scheme; for the proton PDF we have chosen the most recent
parametrization CTEQ6.6M \cite{11} of the CTEQ group.

For the FFs we use the set Belle/CLEO-GM of Ref.\ \cite{12}. Note that
in our previous work \cite{8} we had used the set Global-GM instead,
where also LEP1 data \cite{15} at large $s$ had been included in the
fit, whereas the set Belle/CLEO-GM is based on a fit of the combined
Belle \cite{13} and CLEO \cite{14} data at $\sqrt{s}=10.52$ GeV only.
For the photoproduction cross section $d\sigma/dp_T$ we found results
larger by $25-30\%$ in average when using the Belle/CLEO-GM
parametrization, as compared to the set Global-GM of Ref.\ \cite{12}.
The strong coupling constant $\alpha_s^{(n_f)}(\mu_R)$ is evaluated with
the two-loop formula \cite{16} with $n_f=4$ active quark flavours and
the asymptotic scale parameter $\Lambda^{(4)}_{\overline{\rm MS}} = 328$
MeV, corresponding to $\alpha_s^{(5)}(m_Z)=0.118$. The charm quark mass
is fixed to $m = 1.5$ GeV. We choose the renormalization scale $\mu_R$
and the factorization scales $\mu_F$ and $\mu_{F'}$ related to initial-
and final-state singularities to be $\mu_R = \xi_R m_T$ and $\mu_F =
\mu_{F'} = \xi_F m_T$, where $m_T = \sqrt{m^2+p_T^2}$ is the transverse
mass and $\xi_R$ and $\xi_F$ are parameters varied about their default
values $\xi_R = \xi_F = 1$ in order to assess theoretical uncertainties
as described below.

As already mentioned in the introduction, the photoproduction cross
section is calculated in the GM-VFN scheme. In this scheme the cross
section has a smooth limit for $m \to 0$ and approaches the result of
the ZM-VFN scheme for $m \to 0$ or $p_T \to \infty$. The basic features
of the GM-VFNS are described in Ref.\ \cite{8} and the literature quoted
therein. Compared to the ZM-VFNS, the GM-VFNS incorporates
mass-dependent terms as in the FFNS, so that the cross sections
calculated in the GM-VFNS are supposed to be valid also for medium scale
$p_T$ values, close to the heavy-quark mass. In order to conform with
the $\overline{\rm MS}$ factorization of singularities, finite
subtraction terms must be supplemented to the results of the FFNS
calculation. These subtraction terms had been calculated in Ref.\
\cite{17} for the direct photon, and in Ref.\ \cite{18} for the resolved
photon contributions. They include logarithmic, scale-dependent
contributions related to gluon emission from charm quarks and to
charm-anticharm production from incoming gluons. As a consequence, in
the GM-VFNS one has to take into account also processes with incoming
charm quarks, involving corresponding charm-quark components in the PDFs
of the photon and the proton.  In addition, for the final state, apart
from the FF describing the transition $c \to D^{\ast}$, also FFs for the
transition of a light parton to the heavy meson, $a \to D^{*}$, are
needed. These contributions are not present in the FFNS calculation.
Instead, they are taken into account at fixed order of perturbation
theory as part of the hard scattering cross sections. Compared with the
ZM- and GM-VFNS approaches, the FFNS has the advantage that it is valid
also for $0 < p_T \lsim m$, a property which is not realized in the
presently available implementation of the GM-VFNS.  A reliable
prediction down to $p_T = 0$ is necessary if total cross sections for
heavy-meson production are to be calculated.

In our calculation we implement the experimental conditions of the 
ZEUS analysis: the energies of the incoming protons and electrons 
(positrons) are $E_p = 920$ GeV and $E_e = 27.5$ GeV, respectively. 
The inelasticity $y$ varies in the range $0.02 < y < 0.85$. The 
transverse momentum $p_T$ and the rapidity $\eta$ of the $D^{*}$-meson 
have been measured in the kinematic ranges $1.5 < p_T < 9.0$ GeV and 
$|\eta | < 1.5$. The photon virtuality is taken in the interval $0.05 
< Q^2 < 0.7$ GeV$^2$. 


\section{Results}

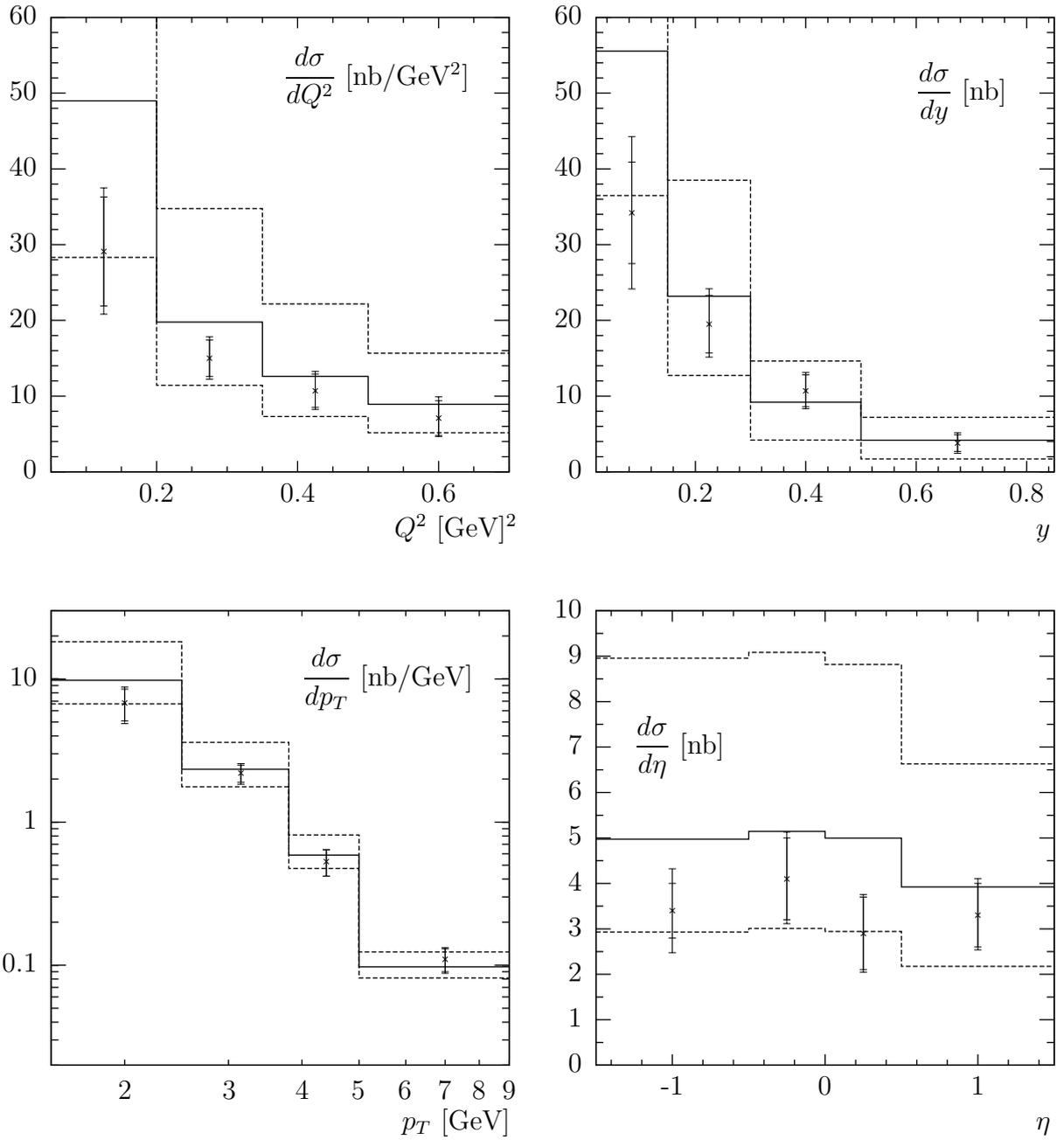
\begin{figure}[ht!] 
\unitlength 1mm
\begin{picture}(160,170)
\put(-23,90){\begin{minipage}[b][80mm][b]{85mm}
          \include{z-fig1a}
          \end{minipage}}
\put(59,90){\begin{minipage}[b][80mm][b]{85mm}
          \include{z-fig1b}
          \end{minipage}}
\put(-24,0){\begin{minipage}[b][80mm][b]{85mm}
          \include{z-fig1c}
          \end{minipage}}
\put(59,0){\begin{minipage}[b][80mm][b]{85mm}
          \include{z-fig1d}
          \end{minipage}}
\end{picture}
\caption{Differential cross sections for $D^{\ast}$-meson production in
  low-$Q^2$ $ep$ scattering, compared with experimental results from the
  ZEUS collaboration \protect\cite{7}. The default choice for the scale
  parameters is $\xi_R = \xi_F = 1$ (full lines) and error bands are
  obtained by varying $\xi_R$ and $\xi_F$ (see text, dashed lines). The
  kinematic range is given by $0.05 < Q^2 < 0.7$ GeV$^2$, $0.02 < y <
  0.85$, $1.5 < p_T < 9.0$ GeV and $|\eta | < 1.5$.}
\label{fig1}
\end{figure}

In this section we present our results for the differential cross
sections of the process $e+p \to e'+D^{*}+X$ in the GM-VFNS as a
function of $Q^2$, $y$, $p_T$ and $\eta$. We choose bins in accordance
with the ZEUS measurement and differential cross sections are obtained
by averaging over the bin sizes as are the experimental ones \cite{7}.
The numerical results are
shown in Fig.\ \ref{fig1}. We have estimated theoretical uncertainties
by varying independently the parameters $\xi_R$ and $\xi_F$ in the range
$0.5 \leq \xi_R$, $\xi_F \leq 2$ about their default values $\xi_R =
\xi_F = 1$ imposing the constraint $0.5 \leq \xi_F/\xi_R \leq 2$. The
maximal and minimal differential cross sections obtained this way are
shown in Fig.\ \ref{fig1} as dashed lines together with the central
values (full lines).
Comparing the errors of the ZEUS data points and the theoretical errors
due to the scale variation we observe that all data points for
$d\sigma/dQ^2$, $d\sigma/dy$, $d\sigma/dp_T$ and $d\sigma/d\eta$ are
compatible with our theoretical predictions. Most of the predictions
with the default scale choice $\xi_R = \xi_F = 1$ are found inside the
range given by the data and their experimental errors. The discrepancies
between theoretical prediction and experimental measurement are largest
for the smallest values of $Q^2$, $p_T$ and $y$; however, also the
theoretical error band is largest for these bins.  Since small values of
these kinematic variables dominate at all values of $\eta$, the
uncertainty due to scale variations is large for all bins of
$d\sigma/d\eta$. For $d\sigma/dp_T$ the relative size of the error band
decreases with increasing $p_T$, as expected. In the first bin of the
$p_T$ distribution where $p_T(\rm min) \simeq 1.5$ GeV, the scale
variable $\mu_F$ is small, approximately equal to $1.5 \times m$, which
explains the large sensitivity to scale variations.

Actually we can not expect our theoretical framework to be particularly
accurate at small $p_T$, close to $p_T \simeq m$, since the cross
section in the GM-VFNS contains parts which are calculated in the
massless approximation. This prevents us to calculate cross sections
down to $p_T = 0$ and consequently also the total $D^{\ast}$-production
cross section integrated over $p_T$ can not be calculated reliably. In
the kinematic region $p_T \lsim m$, the FFNS is the better choice. The
comparison of the data in the first $p_T$ bin with the FMNR result for
photoproduction \cite{2}, as presented in Ref.\ \cite{7}, shows indeed
good agreement. Since $d\sigma/dp_T$ is over-estimated in our approach
for $p_T \simeq m$, also the cross section integrated over the full
range of $p_T$ values considered in the ZEUS experimental analysis is
too large: the measured total cross section quoted in Ref.\ \cite{7}
is\footnote{Statistical and systematic errors, and an error due to the
  uncertainty in the branching ratios, are added in quadrature.}
$\sigma(ep \to e'D^{*}X) = 10.1^{+1.5}_{-1.3}$ nb, whereas the
corresponding theoretical prediction is $13.9^{+10.4}_{-5.8}$ nb. In the
restricted $p_T$-range $2.5 \leq p_T \leq 9$ GeV, we find instead
$4.1^{+2.0}_{-0.9}$ nb to be compared with the experimental value of
$3.9 \pm 0.4$ nb (obtained from the results given in \cite{7} by summing
over the last three $p_T$ bins), i.e.\ for larger $p_T$ the theoretical
cross section in the GM-VFNS agrees with the measured one quite well.

\begin{figure}[ht!] 
\unitlength 1mm
\begin{picture}(160,170)
\put(-23,90){\begin{minipage}[b][80mm][b]{85mm}
          \include{z-fig2a}
          \end{minipage}}
\put(59,90){\begin{minipage}[b][80mm][b]{85mm}
          \include{z-fig2b}
          \end{minipage}}
\put(-24,0){\begin{minipage}[b][80mm][b]{85mm}
          \include{z-fig2c}
          \end{minipage}}
\put(59,0){\begin{minipage}[b][80mm][b]{85mm}
          \include{z-fig2d}
          \end{minipage}}
\end{picture}
\caption{Differential cross sections for $D^{\ast}$-meson production 
in low-$Q^2$ $ep$ scattering in the kinematic range $0.05 < Q^2 < 
0.7$ GeV$^2$, $0.02 < y < 0.85$, $1.5 < p_T < 9.0$ GeV, $|\eta | < 
1.5$. The full cross section (full lines) is separated into its 
direct (dotted lines) and resolved parts (dashed lines) for $\xi_R 
= \xi_F = 1$.}
\label{fig2}
\end{figure}
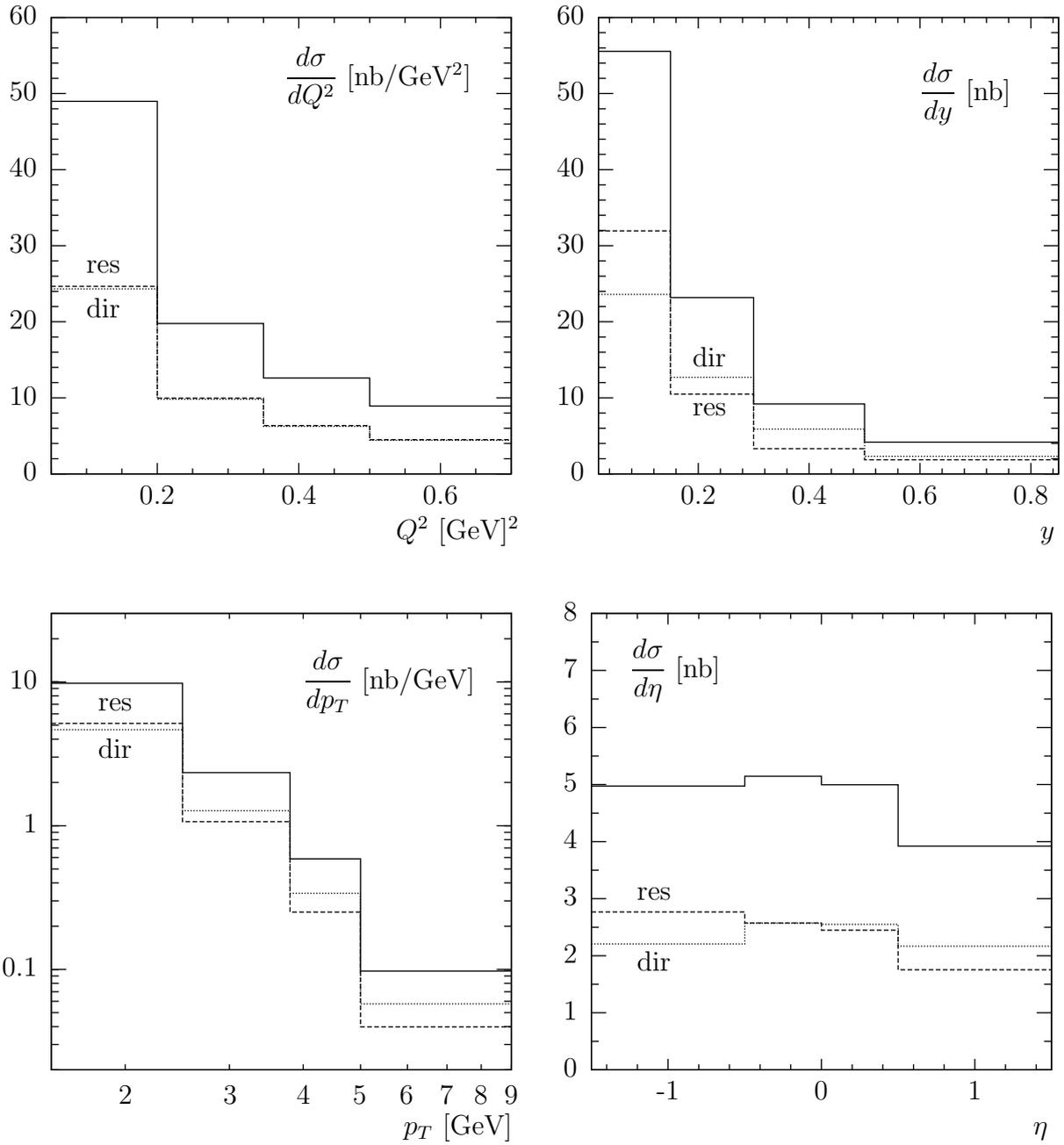

In obtaining theoretical predictions for the photoproduction cross
section, one has to distinguish direct and resolved components.  Each of
the two parts depends on the choice of a factorization scheme and can
not be compared separately to experimental data.  Only the sum of the
direct and resolved contributions has a physical meaning. From the
theoretical point of view, it is nevertheless interesting to study the
decomposition of the cross section into these two components, in
particular in view of a comparison with results obtained in the FFNS
approach for photoproduction \cite{2} or for DIS \cite{1}. In DIS at
large $Q^2$, there is only a direct contribution to the cross section
since initial-state singularities appearing in the limit $Q^2 \to 0$
have not to be subtracted. The photoproduction cross section in the FFN
scheme \cite{2} has a smaller resolved part than in the GM-VFNS, since
it originates only from initial gluons and light quarks, whereas in the
GM-VFNS also processes with charm quarks in the initial state
contribute. Therefore, in our case the direct and resolved contributions
are of comparable magnitude. This is seen in detail in Fig.\ \ref{fig2}
where we have plotted these two components and their sum for
$d\sigma/dQ^2$, $d\sigma/dy$, $d\sigma/dp_T$ and $d\sigma/d\eta$.

We conclude that in the general-mass variable-flavour-number scheme, the
photoproduction approximation, i.e.\ the approximation where the hard
scattering cross sections are evaluated with $Q^2=0$, leads to
theoretical predictions in very good agreement with recent ZEUS data for
finite, but small $Q^2 \neq 0$, in particular for the larger values of
$p_T$. The description of the measurement at the smallest values of $p_T
\simeq 1.5$ GeV is less satisfactory, but still in agreement with data
within errors.


\end{document}